\documentclass[twocolumn]{aastex63}

\usepackage{mathtools}
\usepackage{amsmath}
\usepackage{hyperref}
\usepackage{xcolor}
\usepackage{lipsum}
\usepackage{hyperref}

\shorttitle{Constraints on $B$-mode detection}
\shortauthors{Herv\'ias-Caimapo et al.}

\begin{document}

\title{Galactic foreground constraints on primordial $B$-mode detection for ground-based experiments}

\correspondingauthor{Carlos Herv\'ias-Caimapo}
\email{cherviascaimapo@fsu.edu}

\author[0000-0002-4765-3426]{Carlos Herv\'ias-Caimapo}
\affiliation{Department of Physics, Florida State University, Tallahassee, Florida 32306, USA}
\affiliation{Jodrell Bank Centre for Astrophysics, Department of Physics \& Astronomy, School of Natural Sciences, University of Manchester, Oxford Road, Manchester M13 9PL, U.K.}

\author[0000-0003-4787-2888]{Anna Bonaldi}
\affiliation{SKA Organisation, Lower Withington Macclesfield, Cheshire SK11 9DL, U.K.}

\author[0000-0002-0370-8077]{Michael L. Brown}
\affiliation{Jodrell Bank Centre for Astrophysics, Department of Physics \& Astronomy, School of Natural Sciences, University of Manchester, Oxford Road, Manchester M13 9PL, U.K.}

\author[0000-0001-7109-0099]{Kevin M. Huffenberger}
\affiliation{Department of Physics, Florida State University, Tallahassee, Florida 32306, USA}

\begin{abstract}

Contamination by polarized foregrounds is one of the biggest challenges for future polarized cosmic microwave background (CMB) surveys and the potential detection of primordial $B$-modes. Future experiments, such as Simons Observatory (SO) and CMB-S4, will aim at very deep observations in relatively small ($f_{\rm sky} \sim 0.1$) areas of the sky. In this work, we investigate the forecasted performance, as a function of the survey field location on the sky, for regions over the full sky, balancing between polarized foreground avoidance and foreground component separation modeling needs. To do this, we simulate observations by an SO-like experiment and measure the error bar on the detection of the tensor-to-scalar ratio, $\sigma(r)$, with a pipeline that includes a parametric component separation method, the Correlated Component Analysis, and the use of the Fisher information matrix. We forecast the performance over 192 survey areas covering the full sky and also for optimized low-foreground regions. We find that modeling the spectral energy distribution of foregrounds is the most important factor, and any mismatch will result in residuals and bias in the primordial $B$-modes. At these noise levels, $\sigma(r)$ is not especially sensitive to the level of foreground contamination, provided the survey targets the least-contaminated regions of the sky close to the Galactic poles. 




\end{abstract}


\section{Introduction} \label{sec:intro} 

In the next decade, the potential detection of primordial polarized $B$-modes in the cosmic microwave background (CMB) radiation could be one of the most important milestones in modern cosmology. $B$-modes are a key prediction of inflation. They are predicted to arise due to a background of primordial gravitational waves created just after the Big Bang. The amplitude of these primordial gravitational waves could vary by a few orders of magnitude, from so small that is impossible to detect, to large enough to be detected in the immediate future \citep{2002PhRvL..89a1303K}. The best current limits constrain the tensor-to-scalar ratio to $r_{0.002} < 0.056$ \citep{2018PhRvL.121v1301B,2020A&A...641A..10P}. In practice, the experimental detection of primordial $B$-modes is extremely challenging, and the cosmological community is investing significant effort in experiments specifically designed to meet this challenge, such as satellite experiments like LiteBIRD \citep{2019JLTP..194..443H,2020JLTP..199.1107S}, as well as ground-based telescopes, such as the Simons Observatory \citep[SO; ][]{2019BAAS...51g.147L} and CMB-S4 \citep{2019arXiv190704473A}.

The polarized foregrounds are perhaps the biggest obstacle we need to overcome for such a detection because they block our otherwise unobstructed view of the CMB. The data analysis techniques used to model and mitigate contamination by foregrounds are termed component separation. We have methods to remove most of the contaminating signal, but small residuals left over by systematics and modeling mismatch are a source of bias for the polarization $B$-modes. These residuals could be comparable to the primordial $B$-mode signal we aim to measure. Several works in the recent literature have aimed at forecasting the performance of different satellite and ground-based experiments for measuring the primordial $B$-modes, with an emphasis on the foreground modeling \citep[e.g.][]{betoule_2009,bonaldi_2011,katayama_2011,armitage-caplan_2012,errard_2012,remazeilles_2016,2018JCAP...04..023R,remazeilles_2018,alonso_2017,chluba_2017,hervias_2017,2019arXiv190508888T,2021MNRAS.503.2478R}.

Ground-based experiments, unlike the more expensive satellite experiments, do not cover the full sky. Instead, they must choose an area to target their observations \citep[e.g.][]{2018SPIE10708E..41S,2019arXiv190704473A} and balance between focusing on a small region to increase the map sensitivity and covering the largest possible region to reduce sample variance in the primordial $B$-mode signal (which peaks on large angular scales). Moreover, ideally we would like the complete absence of any foreground contamination when observing the CMB. This is not possible, so ground-based experiments often choose the areas where the foregrounds are the weakest, which in general correspond to areas near the south Galactic pole, observed from the southern hemisphere in Chile and Antarctica. No experiment so far has surveyed the north Galactic pole in its entirety, which may be just as good. However, for measuring primordial $B$-modes, avoiding the foregrounds is certainly not sufficient. Even component separation, which helps in the cleaning of foregrounds, is not sufficient. The selection of an optimal survey area for measuring the primordial $B$-modes (and $r$) needs to consider both factors: on one hand we wish the foregrounds to be as weak as possible (the foreground avoidance approach) while on the other, we require the foregrounds to have sufficient signal-to-noise ratio (S/N) so that component separation algorithms can measure their properties and thus clean them.

In this work, we investigate this issue. We analyze this balance and try to optimize the selection of the survey field in order to obtain the best measurement of the tensor-to-scalar ratio. We use SO as our example survey, but this paper should not be interpreted as an official forecast for this experiment, but rather a guide on the relative performance between different regions of the sky.

In Section~\ref{sec:simulations}, we describe the model and simulated observations we use, with a configuration based on the SO experiment. In Section~\ref{sec:methodology}, we describe our pipeline in detail. In Section~\ref{sec:results}, we present our results for the forecasted error on the detection of the tensor-to-scalar ratio, $\sigma(r)$. In Section~\ref{sec:discussion}, we discuss a few specific issues, regarding the use of extra high-frequency channels to monitor the thermal dust emission. Finally, in Section~\ref{sec:conclusions}, we draw our conclusions.
\section{Simulations} \label{sec:simulations} 

\begin{deluxetable*}{r r r r r r r}
\tablecaption{SO-SAT and CCAT-prime Instrumental specifications Used in the Simulated Observations. \label{table:SO_SAT}}
\tablehead{ 
		\colhead{Band}  & \colhead{Beam FWHM}	& $1/f$ $\ell_{\rm knee}$ & $1/f$ $\alpha_{\rm knee}$ & \colhead{\emph{baseline} sensitivity} & \colhead{\emph{goal} sensitivity} & \colhead{CCAT-prime sensitivity}	\\
		\colhead{[GHz]}  & \colhead{[arcmin]}	& & & \colhead{[$\mu$K$_{\rm CMB} \cdot $arcmin]} & \colhead{[$\mu$K$_{\rm CMB} \cdot $arcmin]} & \colhead{[$\mu$K$_{\rm CMB} \cdot $arcmin]}
}
\startdata
27 & 91 & 15 & -2.4 & 24.71 & 17.65 &- \\
39 & 63 & 15 & -2.4 & 15.29 & 11.76 &- \\
93 & 30 & 25 & -2.5 & 2.83 & 2.00 &-  \\
145 & 17 & 25 & -3.0 & 3.58 & 2.25 &- \\
225 & 11 & 35 & -3.0 &  6.40 & 4.24 & -\\
280 & 9 & 40 & -3.0 & 16.37 & 10.42 &- \\
350 & 0.58 & 700 & -1.4 & - & -& 105\\
410 & 0.5 & 700 & -1.4 & - & -& 372\\
850 & 0.23 & 700 & -1.4 &-&-&$5.7\times10^5$\\
\enddata
\tablecomments{The listed sensitivity is the white-noise part of the curve, while the $1/f$ noise curve parameters follow \citet{2019JCAP...02..056A} for SO and \citet{2020JLTP..199.1089C} for the CCAT-prime instrument.}
\end{deluxetable*}


We simulate observations by modeling the SO Small Aperture Telescopes (SATs) survey as our example experiment, which will target an $f_{\rm sky} \sim 0.1$ area of the sky at degree resolution, with the intention of measuring the primordial $BB$ angular power spectrum and constraining the value of $r$ \citep{2019JCAP...02..056A,2019BAAS...51g.147L}. We refer to the area of this survey as the SO-SAT mask.

The simulated observations of the microwave Galactic foregrounds are created with the \textsc{pysm3} code \citep{thorne_2017}, smoothed with a Gaussian beam of appropriate width in order to model the SO-SAT instrument. The frequency channels, beam widths, and sensitivities used for the simulations are listed in Table~\ref{table:SO_SAT}. The maps are created at $N_{\rm side} = 512$. The model we use includes polarized thermal dust and synchrotron emission. Their spectral energy distribution (SED) is a modified blackBody (MBB) for thermal dust, given by 
\begin{equation} \label{eq:MBB}
    S_{\rm dust}(\nu) \propto \nu^{\beta_{\rm dust}+1} / [\exp(h\nu/kT_{\rm dust})-1]
\end{equation}
in antenna temperature units, where the spectral index $\beta_{\rm dust}$ and the dust temperature $T_{\rm dust}$ are the free parameters, and $h$ and $k$ are the Planck and Boltzmann constant, respectively. For synchrotron, the SED is a power law with an index $\beta_{\rm syn}$ close to 3,
\begin{equation} \label{eq:synch}
    S_{\rm syn}(\nu) \propto \nu^{-\beta_{\rm syn}} \text{.}
\end{equation}
The modeled dust and synchrotron emission correspond to models \textit{d1} and \textit{s1}, respectively, in \citet{thorne_2017}. The synchrotron polarization template is based on the 9 yr WMAP maps \citep{bennett_2013}. The synchrotron SED is the power law from eq.~\ref{eq:synch} with a spatially variable index, which is taken from `model 4' of \citet{miville-deschenes_2008}.

The thermal dust polarization is constructed from the Planck 2015 data release \citep{planck_2015_foregrounds}, using the model fitted with the \textsc{commander} component separation code. The polarization template at the anchor frequency of 353\,GHz, as well as the spatially variable maps of $\beta_{\rm dust}$ and $T_{\rm dust}$ are used to build this dust model.

The CMB realizations are created with a fiducial power spectra created with \textsc{camb} \citep{2012JCAP...04..027H} with the full lensing $BB$ signal, an amplitude of primordial scalar perturbations $A_{\rm s}=2\times10^{-9}$, an index for the primordial scalar perturbation power spectra $n_{\rm s}=0.965$ and $r=0$. Otherwise, we use a standard $\Lambda$CDM cosmology with the Hubble constant $H_0=$\,67.5 km/s/Mpc, baryonic matter density parameter $\Omega_{\rm b} h^2 = 0.022$, cold dark matter density parameter $\Omega_{\rm c} h^2 = 0.122$, sum of neutrino masses $\sum m_{\nu} = 0.06$\,eV, and reionization optical depth $\tau = 0.06$.

We simulate the instrument noise with the public SO noise generator\footnote{\url{https://github.com/simonsobs/so_noise_models}}. This includes the use of atmospheric $1/f$ noise curves with their optimistic $\ell_{\rm knee}$, listed in Table~\ref{table:SO_SAT}. We create two sets of noise realization with the baseline and goal levels. Further details are provided in \citet{2019JCAP...02..056A}. We use only the auto $N_{\ell}$ spectra curves, not the cross-spectra, so the noise is uncorrelated between frequency channels in our simulations. The white-noise level sensitivities, along with the $1/f$ parameters, are listed in Table~\ref{table:SO_SAT}.

We also consider the Cerro Chajnantor Atacama Telescope-prime (CCAT-prime) project in the Fred Young Submillimeter Telescope (FYST) \citep{2020JLTP..199.1089C}, a new submillimeter 6 m telescope scheduled to start observations roughly at the same time as SO. One of the science objectives of the experiment is to contribute to CMB experiments with high-frequency observations and thermal dust monitoring. The synergy between the CCAT-prime project and SO/CMB-S4 will allow for better constraining of the thermal dust foreground that hinders the potential detection of CMB $B$-modes.

We simulate the observations in the 350, 410, and 850\,GHz frequency channels from the Prime-Cam instrument in the FYST, using the same sky model. We use the noise power spectra curves to generate $1/f$ noise simulations as described in \citet{2020JLTP..199.1089C}. The details of the $1/f$ parameters and white-noise levels are listed in Table~\ref{table:SO_SAT}.
\section{Methodology} \label{sec:methodology} 

Our full pipeline consists of running the Correlated Component Analysis (CCA) estimation over a given region of the sky on the simulated observations and subsequently reconstructing the angular power spectra of the three components that are included: CMB, thermal dust, and synchrotron emission. We apply this pipeline to 200 Monte Carlo iterations to estimate the covariance matrix of the $B$-mode angular power spectrum of the CMB. Finally, we estimate a $\sigma(r)$ error using the Fisher information matrix.

\subsection{Component separation and estimation of the $BB$ power spectrum} \label{sec:cca} 

The component separation method used for our analysis is the CCA \citep{bonaldi_2006,ricciardi_2010}. This is a parametric foreground fitting algorithm that exploits a generalized least-squares method applied to second-order statistics. In its harmonic domain version, which we use in this work, it estimates both the frequency behavior of all the foregrounds, expressed as a function of a few parameters, and the auto- and cross-power spectra of the components. A parametric method is particularly useful when investigating the correlation between the goodness of foreground fitting and the intensity of the foregrounds, which is the goal of this analysis.

As with many other component separation techniques, CCA is a linear method. It assumes that the observed signal is a linear combination of the true signal from the different components in the sky plus some noise. By making a few other assumptions (that spectral and spatial features on the components are not correlated, and that the instrumental beam is constant within each passband), the signal can be written for each line of sight $\mathbf{\hat{r}}$ as
\begin{equation} \label{eq:real-space}
    \mathbf{x}(\mathbf{\hat{r}}) = [\mathbf{B} * \mathbf{H s}](\mathbf{\hat{r}}) + \mathbf{n}(\mathbf{\hat{r}}) \text{,}
\end{equation}
where $\mathbf{x}$ and $\mathbf{n}$ are vectors containing the observed signal and noise, respectively, $\mathbf{B}$ is a matrix containing the beam per frequency channel, $\mathbf{s}$ is the vector containing the true signal of each component, and the symbol $*$ denotes convolution. $\mathbf{H}$ is the mixing matrix that contains the frequency spectra (SED) of all the components (three in our case: CMB, synchrotron, and thermal dust). If we transform to harmonic space, eq.~\ref{eq:real-space} becomes
\begin{equation}
    \tilde{\mathbf{x}} = \tilde{\mathbf{B}} \mathbf{H} \tilde{\mathbf{s}} + \tilde{\mathbf{n}} \text{,}
\end{equation}
where $\tilde{\mathbf{x}}$, $\tilde{\mathbf{s}}$, and $\tilde{\mathbf{n}}$ are the harmonic transforms of $\mathbf{x}$, $\mathbf{s}$ and $\mathbf{n}$, respectively, and  $\tilde{\mathbf{B}}$ is the harmonic transform of $\mathbf{B}$.

The relation between the covariances is given by
\begin{equation} \label{eq:second-order-stats}
    \tilde{\mathbf{C}}_{\mathbf{x}} = \tilde{\mathbf{B}} \mathbf{H} \tilde{\mathbf{C}}_{\mathbf{s}} \mathbf{H}^{T} \tilde{\mathbf{B}}^{\dagger} + \tilde{\mathbf{C}}_{\mathbf{n}} \text{,}
\end{equation}
where the dagger superscript denotes the adjoint matrix, and the angular power spectra $\mathbf{C}$ for the observations $\mathbf{x}$, for the true signal $\mathbf{s}$ and noise $\mathbf{n}$. $\mathbf{C}_{\mathbf{x}}$ and $\mathbf{C}_{\mathbf{n}}$ are estimated from the data and noise properties, $\tilde{\mathbf{B}}$ is also known, while $\mathbf{C}_{\mathbf{s}}$ and $\mathbf{H}$ are the unknowns.

To reduce the number of unknowns, the mixing matrix is expressed as a function of a few parameters $\mathbf{H}=\mathbf{H}(\mathbf{p})$. We use the blackbody spectrum for the CMB (with no free parameters); eq.~\ref{eq:synch} for  synchrotron, with the spectral index $\beta_{\rm syn}$ as a free parameter; and eq.~\ref{eq:MBB} for thermal dust, where the spectral index $\beta_{\rm dust}$ and the temperature $T_{\rm dust}$ are, in general, both free parameters.

Finally, the data and source spectra are binned in multipoles to further reduce the number of unknowns. Then, in harmonic space, our model is the following equation:
\begin{equation} \label{eq:dV}
    \mathbf{d}_V = \mathbf{\mathcal{H}}_{kB} \mathbf{c}_V + \epsilon_{V} \text{,}
\end{equation}
where $\mathbf{d}_V$ are the noise-bias-subtracted auto- and cross-power spectra of the observations and arranged into a vector over the multipole bins $\hat{\ell}$.  $\mathbf{\mathcal{H}}_{kB}$ is a block-diagonal matrix that contains one block per multipole bin. Each block is a matrix given by $\mathbf{\mathcal{H}}_{k}(\hat{\ell}) = [\tilde{\mathbf{B}}(\hat{\ell}) \mathbf{H}] \otimes [\tilde{\mathbf{B}}(\hat{\ell}) \mathbf{H}] $, where $\otimes$ is the Kronecker product (hence the $k$ subindex). This matrix, just like the mixing matrix, depends on the parameters $\mathbf{p}$. $\mathbf{c}_V$ is the power spectra of the different source components in the model, again arranged into a vector over multipole bins, and $\epsilon_{V}$ represents the residuals of the power spectra, between the true sources in the sky and the modeled source components.

\citet{ricciardi_2010} shows that the mixing matrix parameters, $\mathbf{p}$, and the binned power spectra of the components, $\mathbf{c}_V$, can be found by minimizing the functional
\begin{equation} \label{eq:functional}
    \mathbf{\Phi}(\mathbf{p},\mathbf{c}_V) = [ \mathbf{d}_V - \mathbf{\mathcal{H}}_{kB}(\mathbf{p}) \cdot \mathbf{c}_V ]^T  \mathbf{\mathcal{N}}_{\epsilon B}^{-1}  [ \mathbf{d}_V - \mathbf{\mathcal{H}}_{kB}(\mathbf{p}) \cdot \mathbf{c}_V ] \text{,}
\end{equation}
where $\mathbf{\mathcal{N}}_{\epsilon B}$ is a block-diagonal matrix, containing one block per multipole bin. Each block is the matrix $\mathbf{N}_{\epsilon}(\hat{\ell})$, which corresponds to the auto and cross terms (over the frequency channels) of the Gaussian covariance matrix for the noise power spectra.

For each value of the mixing matrix parameters $\mathbf{p}$, the estimated source angular power spectra $\bar{\mathbf{c}}_V$ are obtained as a suitable linear mixture of the data power spectra, depending on the estimated mixing matrix $\mathbf{\mathcal{H}}_{kB}(\mathbf{p})$ and the noise covariance matrix $\mathbf{\mathcal{N}}_{\epsilon B}$, given by
\begin{equation} \label{eq:weights}
    \bar{\mathbf{c}}_V(\mathbf{p}) = [ \mathbf{\mathcal{H}}_{kB}^T(\mathbf{p}) \mathbf{\mathcal{N}}_{\epsilon B}^{-1} \mathbf{\mathcal{H}}_{kB}(\mathbf{p}) ]^{-1} \mathbf{\mathcal{H}}_{kB}^T(\mathbf{p}) \mathbf{\mathcal{N}}_{\epsilon B}^{-1} \mathbf{d}_V \text{.}
\end{equation}
The solution for both the mixing matrix and the source spectra yields the minimum value of eq.~\ref{eq:functional}. The minimization in CCA is performed with a simulated annealing over the foreground spectral parameters. 

We have modified the CCA method slightly as compared to the one presented in \citet{ricciardi_2010}. The estimation is done over the polarization spectra $EE$ and $BB$, as opposed to working over the spectrum of any random scalar field, as before. We also used a more general noise covariance $\mathbf{N}_{\epsilon}(\hat{\ell})$ than the white-noise-only approximation used in \citet{ricciardi_2010}. The \textsc{namaster} software \citep{2019MNRAS.484.4127A} has been used to estimate the angular power spectra of the data and the noise as well as the noise covariance matrix (whose estimation adapts the analytical estimate for a Gaussian covariance proposed in \citet{2004MNRAS.349..603E,2017A&A...602A..41C}). We set the cross-power spectra to zero because we do not simulate noise cross-correlation among the frequency channels.

\subsection{Estimation of the $r$ uncertainty} \label{sec:fisher}

To translate the error on the $BB$ power spectrum to an uncertainty on $r$, $\sigma(r)$, we use the Fisher information matrix. This is given as
\begin{equation}
    F_{ij} = \sum_{\hat{\ell},\hat{\ell}'=1}^{N_b} \mathbf{\mathcal{C}}_{\hat{\ell},\hat{\ell}'}^{-1} \frac{\partial c_{\hat{\ell}}}{\partial p_i} \frac{\partial c_{\hat{\ell}'}}{\partial p_j}  \text{,}
\end{equation}
where $N_b$ is the number of multipole bins, $\mathbf{\mathcal{C}}$ is the covariance matrix over multipole bins of the observable, and $\partial c_{\hat{\ell}} / \partial p_i$ is the binned derivative of the fiducial model spectra with respect to the parameter $p_i$. The covariance matrix of the parameters is given by the inverse of $F_{ij}$. Therefore, the uncertainty of parameter $i$ (marginalized over the rest of the parameters) is given by the square root of the $ii$ element of the inverse of the Fisher information matrix. The off-diagonal elements will measure the correlation between parameters $i$ and $j$.

In our case, the observable is the estimated CMB $BB$ power spectrum $C_{\ell}^{BB}$ and the parameter we are interested in is the tensor to scalar ratio $r$. To correct for residual foreground contamination in the CMB $BB$ power spectrum, we introduce a second parameter, $A_{\rm fore}$, which represents the residual foreground amplitude. We do not consider the foreground spectral parameters explicitly in the Fisher matrix, but they are included in the estimation of the $\mathbf{\mathcal{C}}_{\hat{\ell},\hat{\ell}}$ covariance via the Monte Carlo simulation. Therefore, our fiducial model for the $BB$ power spectrum is the following:
\begin{equation} \label{eq:model}
    \begin{split}
        c_{\ell}(p) = C_{\ell}^{BB, \rm fiducial}(r,A_{\rm fore}) = r C_{\ell}^{BB, \rm primordial}(r=1) \\+ C_{\ell}^{BB, \rm lensing} + A_{\rm fore} C_{\ell}^{BB, \rm foregrounds}(p) \text{.}
    \end{split}
\end{equation}
The covariance matrix of the observable $\mathbf{\mathcal{C}}$ is easily calculated from 200 Monte Carlo iterations over the simulations. The derivative of the fiducial $BB$ power spectra with respect to $r$ and $A_{\rm fore}$ is calculated numerically with respect to mean values $r=0$ and $A_{\rm fore}=1$.

To estimate $C_{\ell}^{BB, \rm foregrounds}$, we produced companion simulations with only the foregrounds, without CMB and noise. This dataset is mixed together with the same weights calculated by CCA, given in eq.~\ref{eq:weights}. Finally, the power spectra $C_{\ell}^{BB, \rm foregrounds}$ is given by the average over the 200 Monte Carlo simulations. This is labeled method A. We realize this procedure is not feasible with real observations, because pure foregrounds maps are not available. We also consider a second method, which is very pessimistic: Because thermal dust is the dominant foreground (as will be shown in Sec.~\ref{sec:using-high-frequency-channels}), we produce 200 Gaussian realizations following a power-law angular power spectrum fitted in \citet{planck_2018_xi} using a mask labeled Large Region (LR) 71, with $f_{\rm sky} = 0.71$. We extrapolate this Gaussian realization map at 353\,GHz to other frequencies using a standard MBB (eq.~\ref{eq:MBB}) with the $\beta_{\rm dust}$ and $T_{\rm dust}$ maps from \citet{planck_intermediate_48}. Note that because we are using the LR71 mask including the full Galactic emission, power will be grossly overestimated in the cleanest regions, but because we are using the Fisher information matrix, the derivative will cancel the absolute amplitude of the foreground residual spectrum. We mix the 200 power-law spectra Gaussian realizations with the same weights calculated by CCA, as described above. This is labeled method B. These two methods are the extremes in realism when accounting for foreground residuals in a full pipeline for the posterior of $r$, but in real observations, the data analysis will be closer to method A than B, because we could use foreground models like the ones from \textsc{pysm}, which is significantly more realistic than a Gaussian random field. We calculate all of our power spectra using $D_{\ell} = \ell(\ell+1)C_{\ell}/2\pi $ instead of $C_{\ell}$.
\section{Results} \label{sec:results} 

\begin{figure*}
    \centering
    \includegraphics[width=1.0\textwidth]{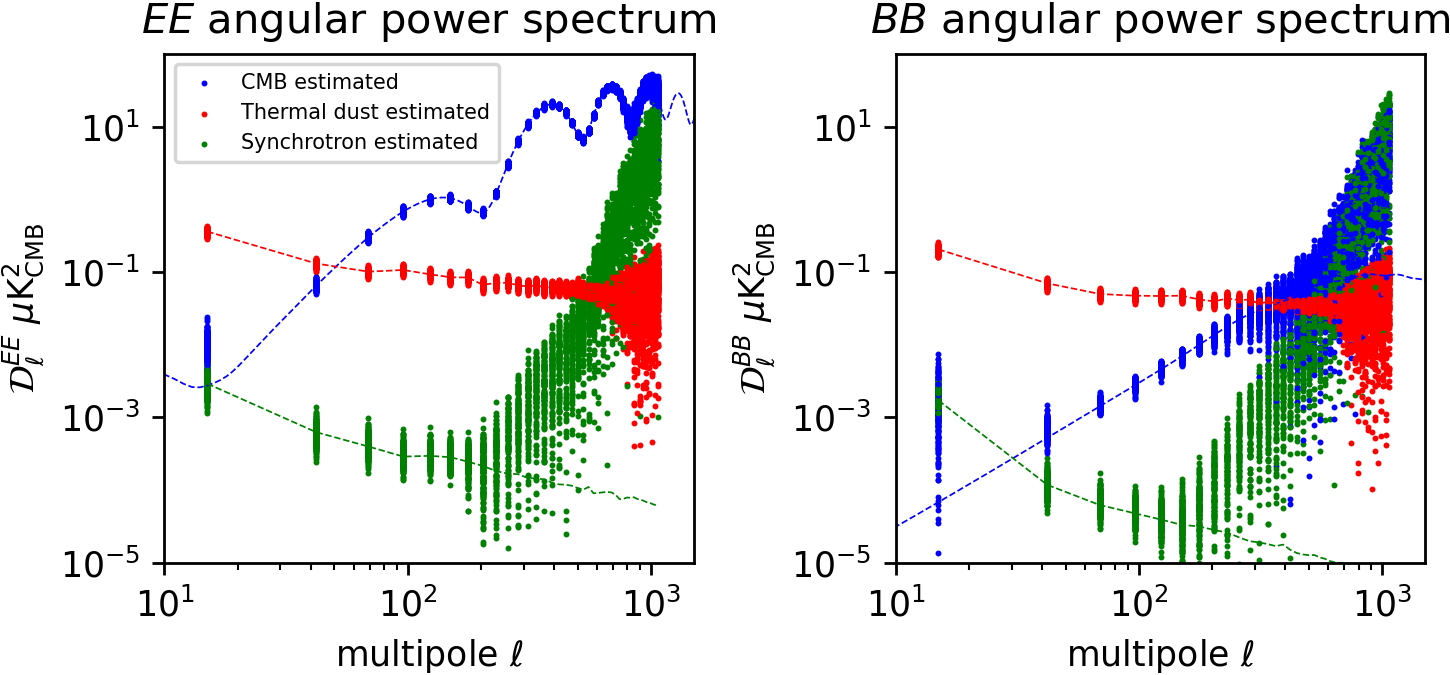}
    \caption{Example reconstructed angular power spectra for the SO-SAT mask (footprint shown in Fig.~\ref{fig:sigmar_map}) with six SO-SAT channels in the baseline noise level. The left is the $EE$ power spectrum, and the right is the $BB$ power spectrum. The dashed blue, green, and red lines are the fiducial spectra for the CMB, synchrotron, and thermal dust emission, respectively. The simulated CMB has $r=0$. The points with the same color scheme correspond to 200 Monte Carlo simulations of the reconstructed spectra. The reference frequency for the foregrounds is 145\,GHz. Note that there is no significant bias in the estimation of the CMB $BB$ spectrum (blue points in the right-hand-side plot).
    }
    \label{fig:so_sat_mask}
\end{figure*}

To represent the SO-SAT instrument specifications, we consider a survey over a disk-shaped region covering $f_{\rm sky} = 0.1$, which is a disk with a radius of $36.8^{\circ}$. These masks are apodized by smoothing with a $1^{\circ}$ FWHM Gaussian kernel. We define 192 of such regions distributed over the full sky, with centers matching the pixels in a \textsc{healpix} pixelization with $N_{\rm side} = 4$. In each of these regions, we run our pipeline to get the uncertainties on $r$ through component separation, power spectrum estimation, and the Fisher matrix analysis.

As mentioned in Sec.~\ref{sec:methodology}, the thermal dust is modeled in the mixing matrix with eq.~\ref{eq:MBB}, with the spectral index $\beta_{\rm dust}$ and the temperature $T_{\rm dust}$ potentially both free parameters. However, because the SO-SAT has a limited range of observing frequencies ($\nu \leq 280$\,GHz) where thermal dust dominates, constraining the two thermal dust spectral parameters ($\beta_{\rm dust}$ and $T_{\rm dust}$ in eq.~\ref{eq:MBB}) at the same time is very difficult. A perturbation of $\pm 1$\,K in the temperature of a fiducial MBB with $\beta_{\rm dust}=1.53$, $T_{\rm dust}=23$\,K, and an anchor frequency of 27\,GHz, results in a change of $\sim 1$\% in the SED at 280\,GHz. To be able to constrain $T_{\rm dust}$ precisely, more frequency channels are required in the range where $T_{\rm dust}$ affects the behavior of the MBB, $\nu > 300$\,GHz. We therefore find that for the fiducial six frequency channels SO-SAT configuration we need to fix the value of $T_{\rm dust}$ to 19\,K, a representative value of the dust temperature away from the Galactic plane region, and leave only the dust spectral index $\beta_{\rm dust}$ as a free parameter.

As a first example, Fig.~\ref{fig:so_sat_mask} shows the beam- and mask-corrected reconstructed binned power spectra of the three sources for the SO-SAT mask in the fiducial six channels with baseline noise configuration. The dashed lines show the fiducial power spectra for the three components in the model, calculated for the two foregrounds by running the power spectrum estimation in the considered region over the simulated model with only synchrotron or only thermal dust emission, respectively. The individual points represent the reconstructed angular power spectra for each source for each of the 200 Monte Carlo simulations.

We note the discrepancies between the model spectra and the estimated spectra at high multipoles, clearly visible in the synchrotron and to a lesser extent in the CMB. This arises from the fact that the channels that constrain synchrotron are mostly the 27 and 39\,GHz channels, which have low spatial resolution. Therefore, at small angular scales, the synchrotron reconstruction is very noisy. The difficulty in accurately measuring the polarized foregrounds in this region, because they are weaker, has an impact on the scatter of the reconstructed spectra for the thermal dust and synchrotron. This large scatter on the foregrounds spectra also somewhat impacts the CMB reconstruction.

\begin{figure*}
    \centering
    \includegraphics[width=1.0\textwidth]{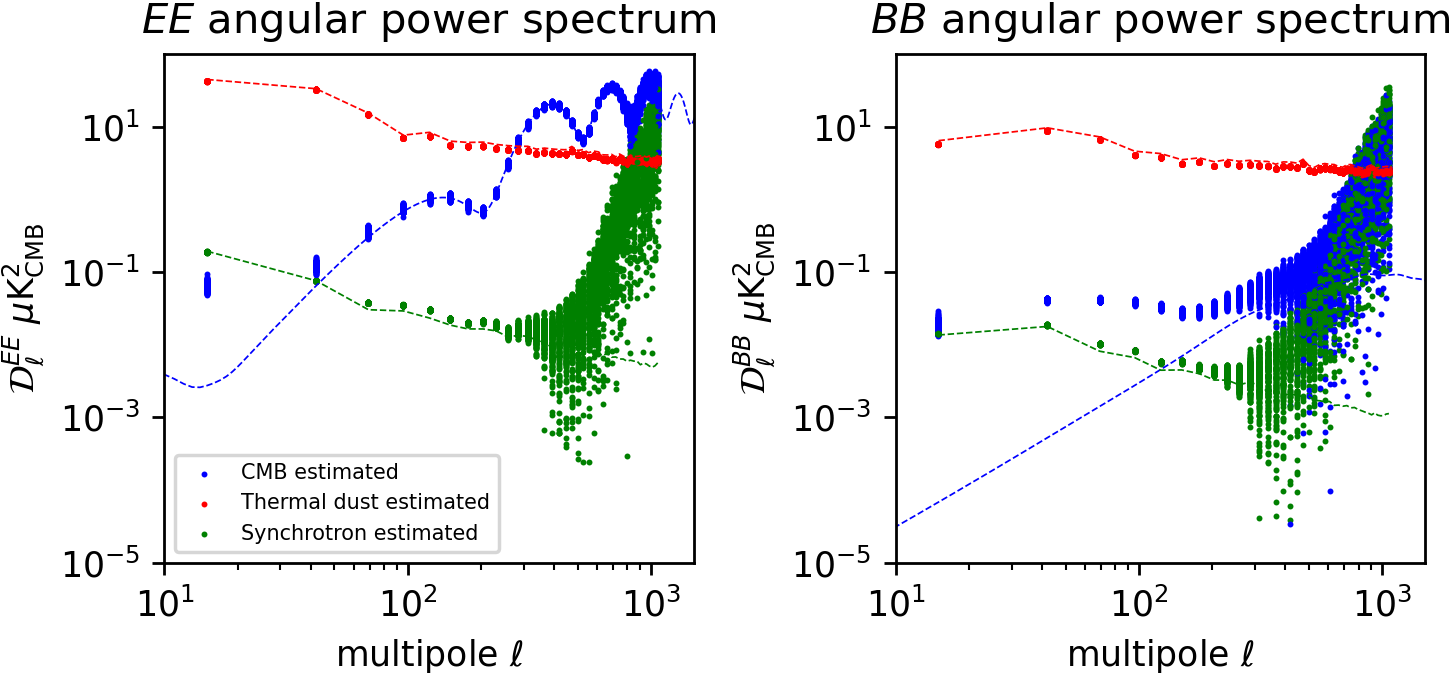}
    \caption{Analogous to Fig.~\ref{fig:so_sat_mask}. Reconstructed power spectra for the high-foreground masked region centered at Galactic coordinates $l=0.0^{\circ}$, $b=9.6^{\circ}$, close to the GC. The reference frequency is 145\,GHz. The simulated CMB has $r=0$. Note the substantial bias due to foreground contamination in the estimation of the CMB $BB$ spectrum (blue points in the right-hand-side plot).
    }
    \label{fig:example_reconstructed_spectra}
\end{figure*}

As another example, we consider in Fig.~\ref{fig:example_reconstructed_spectra} the case for the masked region centered at Galactic coordinates $l=0.0^{\circ}$, $b=9.6^{\circ}$, very close to the Galactic Center (GC). In this case, there is a sizeable bias seen in the CMB $BB$ estimated spectrum, the product of the large impact of polarized foreground residual in the GC region. The scatter on the spectra reconstruction of the foreground components, however, is smaller compared with the SO-SAT mask case. The comparison between Figs.~\ref{fig:so_sat_mask} and \ref{fig:example_reconstructed_spectra} nicely illustrates the kind of trade-off between foreground bias and foreground scatter that this work aims to investigate.

\begin{deluxetable*}{cccccc}
    \tablecaption{Results for the $\sigma(r)$ forecast in units of $10^{-3}$, both in the SO-SAT mask as well as the best result from the 192 masked regions over the full sky, for both the somewhat optimistic method A and the Very pessimistic method B \label{table:results}}
    \tablehead{ \colhead{Instrumental} \vspace{-0.2cm} & \colhead{$\sigma(r)$ SO-SAT} & \colhead{$\sigma(r)$ SO-SAT}	& \colhead{$\sigma(r)$ Best} & \colhead{$\sigma(r)$ Best} & \colhead{Center of Best}\\
    \colhead{Configuration}  & \colhead{Mask A} & \colhead{Mask B}	& \colhead{Masked Region A} & \colhead{Masked Region B} & \colhead{Masked Region}
    }
    \startdata
    SO-SAT fiducial, baseline noise & $4.9$ & $5.2$ & $3.9$ & $6.8$ & $l=225.0\arcdeg$ $b=-78.3\arcdeg$ \\
    SO-SAT fiducial, goal noise & $3.3$ & $3.4$ & $3.1$ & $4.9$ & $l=45.0\arcdeg$ $b=-78.3\arcdeg$ \\
    SO-SAT extended, goal noise & $3.1$ & $3.5$ & $3.0$ & $4.9$ & $l=45.0\arcdeg$ $b=-78.3\arcdeg$ \\
    \enddata
    \tablecomments{Fiducial corresponds to the six SO-SAT frequency channels. Extended corresponds to the fiducial plus 3 extra high frequency CCAT-p channels.}
\end{deluxetable*}

\subsection{Results for $\sigma(r)$ } \label{sec:results_for_sigmar}

\begin{figure*}
    \centering
    \includegraphics[width=0.9\textwidth]{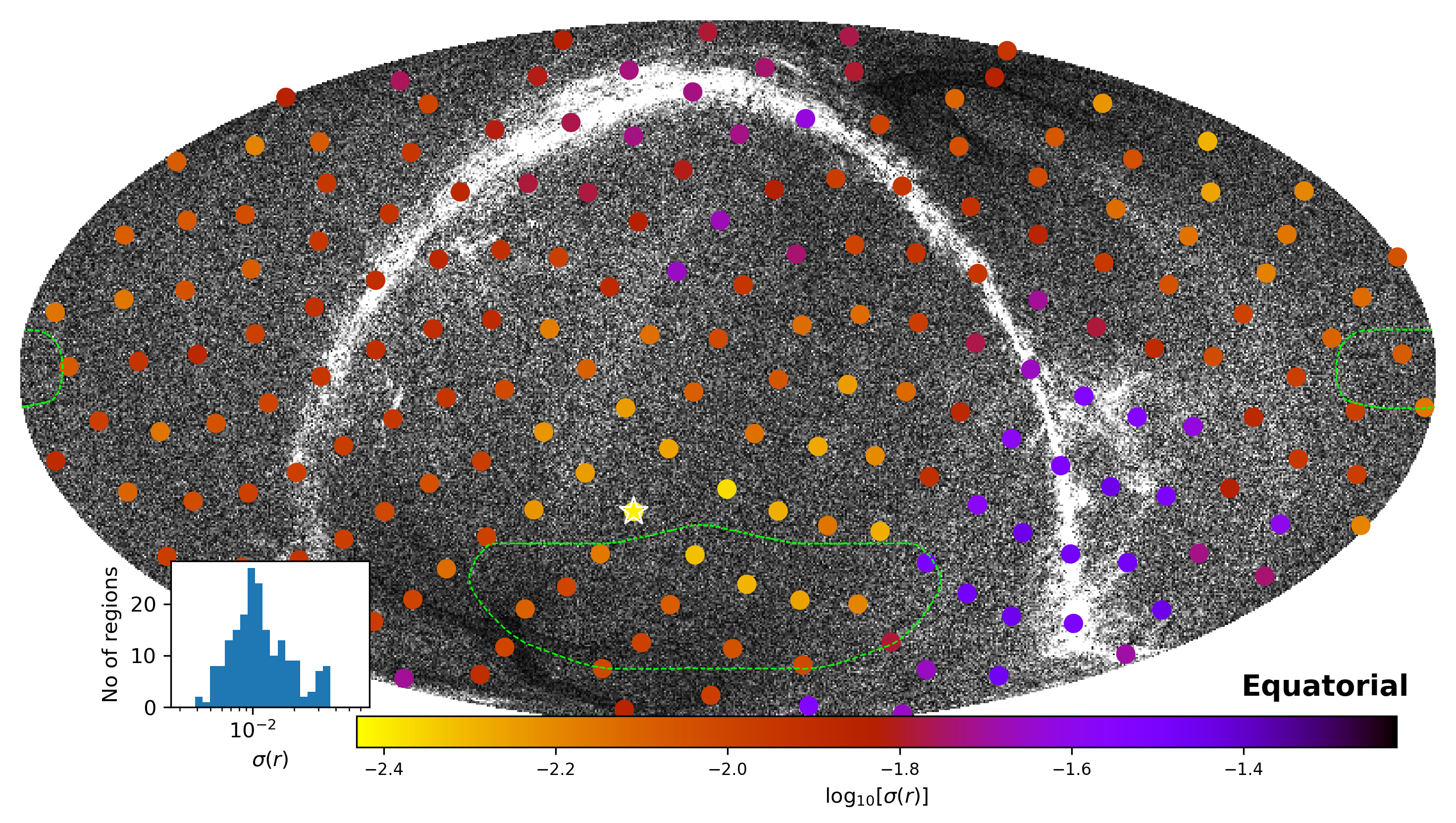}
    \caption{$\sigma(r)$ estimate for method A and 192 masked regions for the SO-SAT fiducial baseline noise configuration. The color of the point centered on the region represents $\sigma(r)$. The background map is the polarization intensity $P$ for Planck 353\,GHz. The dashed light-green contour corresponds to the SO-SAT mask footprint. The star shows the location of the region with the best result. These are the $\sigma(r)$ marginalized over $A_{\rm fore}$. The histogram in the lower-left corner shows the distribution of $\sigma(r)$ from each masked region.}
    \label{fig:sigmar_map}
\end{figure*}

\begin{figure}
    \centering
    \includegraphics[width=1.0\columnwidth]{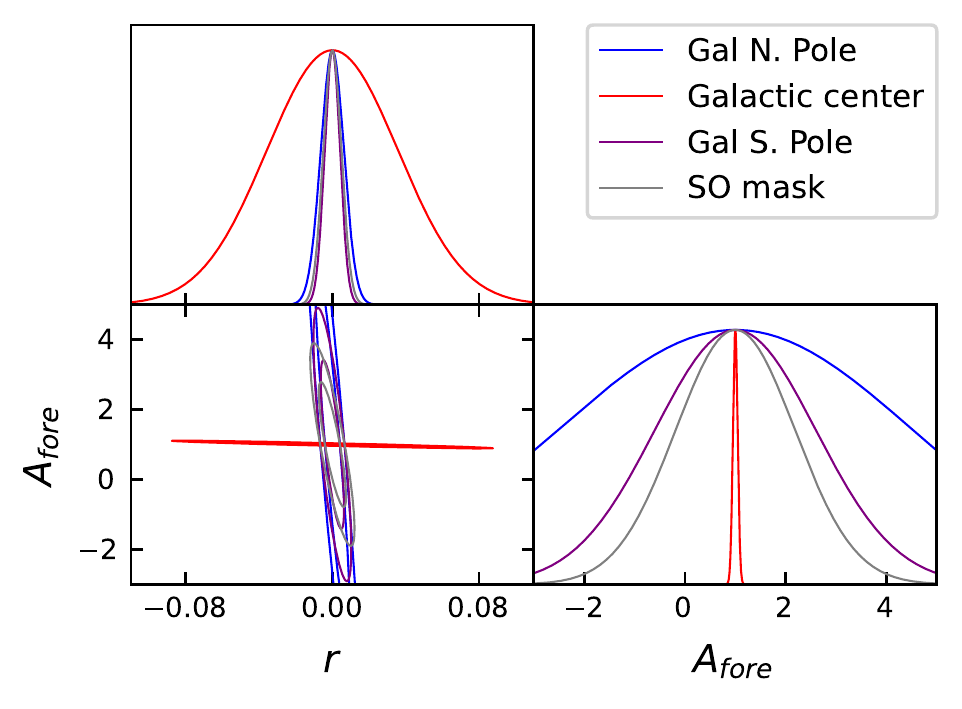}
    \includegraphics[width=1.0\columnwidth]{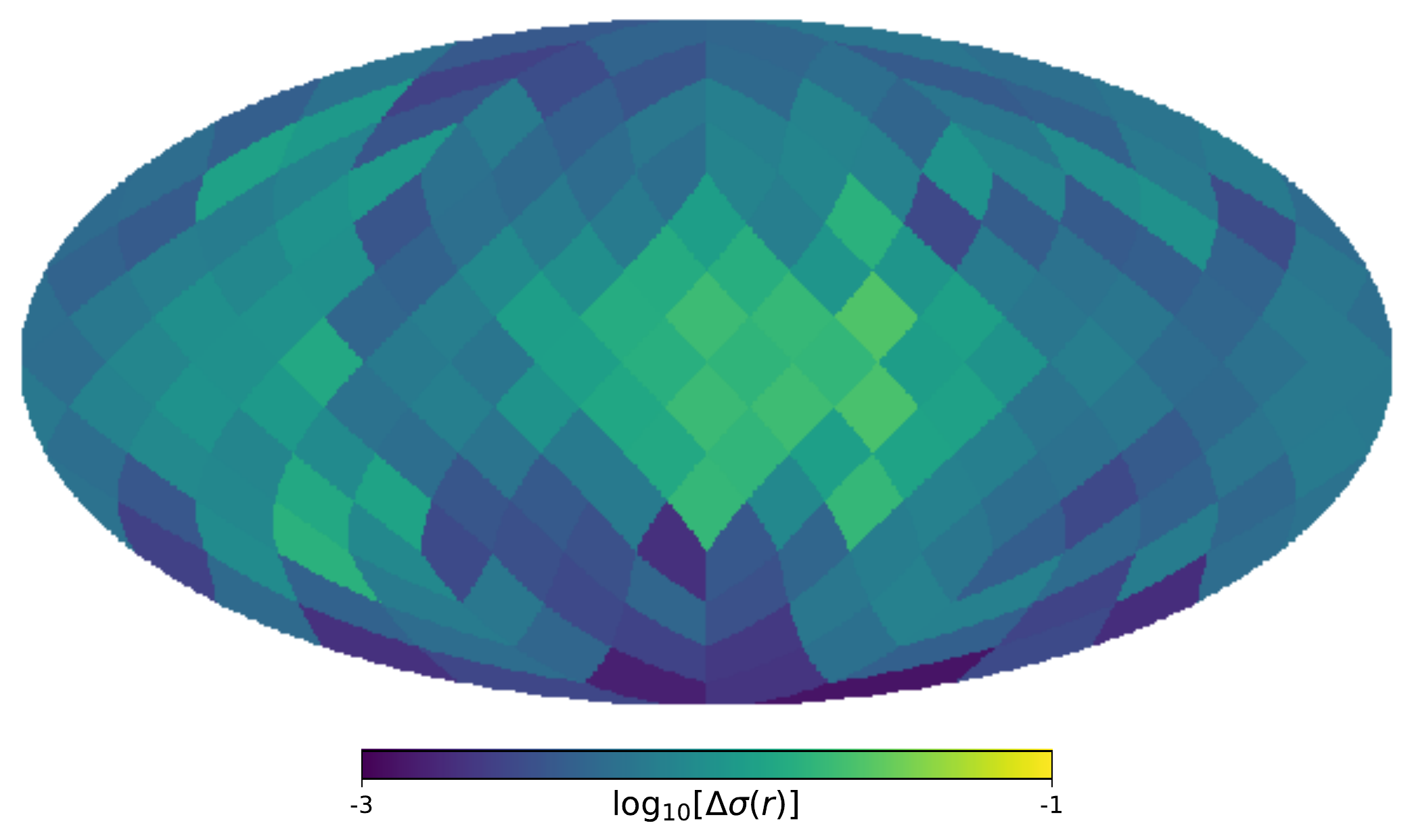}
    \caption{Top: constraints on the $r$ and $A_{\rm fore}$ parameters for the baseline noise level using method A. 1$\sigma$ and 2$\sigma$ contours, along with the marginalized posterior probability for each parameter, for four representative masked regions: one close to the Galactic north pole, one close to the Galactic south pole, one close to the GC, and the SO-SAT mask. Bottom: $\Delta \sigma(r)$ over all 192 masked regions in Galactic coordinates calculated with method A. $\Delta \sigma(r)$ is the difference between the $\sigma(r)$ marginalized over $A_{\rm fore}$ and the $\sigma(r)$ without marginalization.}
    \label{fig:contours}
\end{figure}

Our estimation of $\sigma(r)$ (marginalized over $A_{\rm fore}$ for method A, using the pure foregrounds to construct the foreground residual spectrum) over the 192 $f_{\rm sky}=0.1$ circular regions is shown in Fig.~\ref{fig:sigmar_map}, for the SO-SAT fiducial baseline noise configuration. Each dot is located at the center of the region and its color represents the estimated $\sigma(r)$ value according to the color scale. In the background, we plot the polarization intensity $P=\sqrt{Q^2+U^2}$ for the Planck 353\,GHz frequency map \citep{2020A&A...641A...3P}. The light-green contours show the SO-SAT mask footprint. We also show the histogram of $\sigma(r)$ over the masked regions in the lower-left corner.

In general, the $\sigma(r)$ values are of the order of a few $10^{-2}$ for regions in the Galactic plane, due to foreground contamination. The regions with the lowest $\sigma(r)$ are close to the north and south Galactic poles. Most of them coincide with the location of the SO-SAT mask (shown as green contours in Fig.~\ref{fig:sigmar_map}), with some few exceptions of regions with high decl., not observable from the SO site in the Atacama desert in Chile, above the SO-SAT mask contours at the right and left edges. The estimation over the SO-SAT mask yields $\sigma(r) = 4.9 \times 10^{-3}$ ($2.7 \times 10^{-3}$ before marginalization over $A_{\rm fore}$).

Table~\ref{table:results} lists the $\sigma(r)$ constraints for both the SO-SAT mask and the best result (for method A) of the 192 masked regions considered in this work, as well as the result for method B for the same masks, for the SO-SAT fiducial baseline noise configuration.

The $\sigma(r)$ estimated in the forecast done by SO \citep{2019JCAP...02..056A} finds $\sigma(r) = 1.9 \times 10^{-3}$ in the Fisher matrix configuration. Our result is $\sim 2.5$ times that value. We can attribute this difference to several factors. First, the SO Fisher forecast is a different method from what we do in this work. Their approach is a direct likelihood Monte Carlo Markov Chain (MCMC) of the simulated observations' auto- and cross-spectra, modeling the foregrounds and the CMB at the same time, skipping component separation, an approach followed in, e.g. \citet{2015PhRvL.114j1301B}. The foregrounds are modeled as power laws in multipole space, as well as an MBB for thermal dust and a power law for synchrotron in frequency space. Having the former constraint as a stronger prior would lead to improved agreement between the simulated and (likelihood) modeled foreground components, and in turn to an improved estimate of the CMB signal. This is a constraint we do not impose in our component separation modeling (we only impose the SEDs in frequency space). Also, the fact that we perform component separation in a previous and separate step does amplify the noise to some level when inverting matrices at the point of reconstructing the power spectra of the components, which would explain some of the increased scatter of the $D_{\ell}^{BB}$ estimates, which translate into a larger $\sigma(r)$ in the Fisher forecasts. Another difference between our modeling and the one done in \citet{2019JCAP...02..056A} is that the Fisher matrix of the $D_{\ell}^{BB}$ is calculated as the second derivatives of the likelihood around the fiducial parameters, while in our case this is calculated empirically over 200 Monte Carlo iterations. Finally, the effect of the $A_{\rm fore}$ parameter does increase $\sigma(r)$ in our approach when marginalizing over it, as it does to other component separation methods in \citet{2019JCAP...02..056A} that do marginalize over foreground residuals, such as xForecast, BFoRe and ILC. These methods see an increment in the error bar to $\sigma(r) \sim 3.4 \times 10^{-3}$.

In Fig.~\ref{fig:contours} (top), we show some examples of the correlation between $r$ and $A_{\rm fore}$ for method A. $A_{\rm fore}$ is better constrained in regions with high foreground contamination. At the same time, $r$ and $A_{\rm fore}$ are anticorrelated in these same regions and that anticorrelation diminishes toward zero as we move away from the GC towards the foreground-free Galactic pole regions. Fig.~\ref{fig:contours} (top) shows this for four representative regions: one close to the north Galactic pole, one close to the south Galactic pole, one close to the GC, and the SO-SAT mask. This correlation allows us to quantify the bias introduced by the foreground contamination.

In Fig.~\ref{fig:contours} (bottom) we compare the $\sigma(r)$ constraint in all 192 circular masked regions with and without marginalization over the $A_{\rm fore}$ parameter for method A \footnote{The former is calculated as $\sqrt{ \left[ F^{-1} \right]_{00}}$, with $F$ being the Fisher matrix and the index 0 corresponding to the parameter $r$. The latter is calculated as $\sqrt{1/F_{00}}$}. $\Delta \sigma(r)$ is the difference between the marginalized and unmarginalized $\sigma(r)$ error bar. We can see that the foreground contaminated regions along the Galactic plane have a much larger $\Delta \sigma(r)$. Weak polarized foreground regions in the Galactic poles have $\Delta \sigma(r)$ of a few $10^{-3}$, while strong polarized foreground regions in the Galactic plane have $\Delta \sigma(r)$ of a few $10^{-2}$. In practice, this shows the effect of marginalizing over $A_{\rm fore}$ and how it leads to a $\sigma(r)$ parameter that is reflective of both scatter as well as bias due to strong foreground contamination.

\subsection{Different instrumental configurations} \label{sec:different-instruments} 

\begin{figure}
    \centering
    \includegraphics[width=1.0\columnwidth]{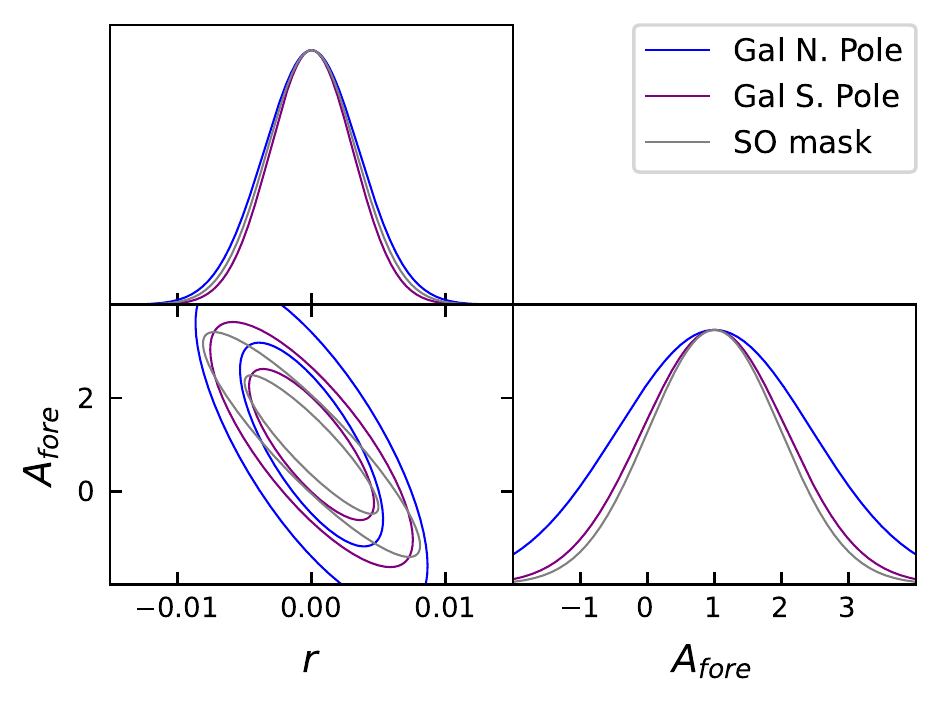}
    \includegraphics[width=1.0\columnwidth]{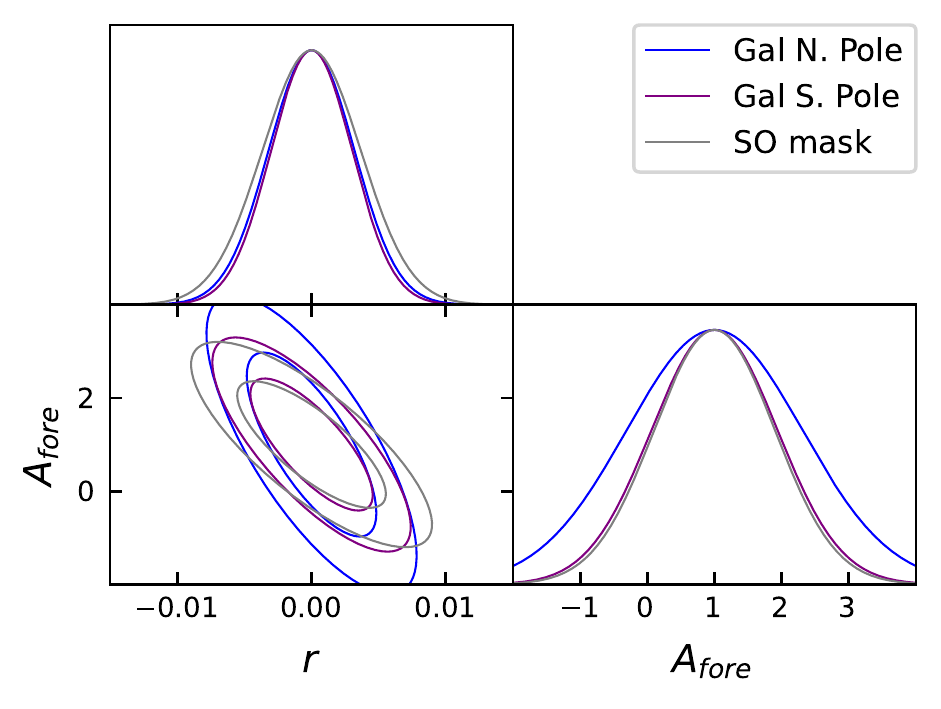}
    \caption{Confidence contours and posterior probabilities for $r$ and $A_{\rm fore}$ for method A and for three representative masked regions. Top: fiducial SO-SAT channels with goal noise levels. Bottom: the same for the extended configuration, the fiducial SO-SAT channels with goal noise levels supplemented by three extra CCAT-prime high-frequency channels.}
    \label{fig:contours_extra_configs}
\end{figure}

Besides the fiducial SO-SAT instrumental configuration with baseline noise level, we consider two extra instrumental configurations: the fiducial SO-SAT channels but with the lower noise goal performance; and the fiducial SO-SAT channels with the goal noise performance supplemented with the three CCAT-prime channels at 350, 410, and 850\,GHz, labeled the extended configuration. The results for these configurations are listed in Table~\ref{table:results}.

The improvement introduced by the lower noise level is evident. The $\sigma(r)$ constraint over the SO-SAT mask improves from $\sigma(r) = 4.9 \times 10^{-3}$ to $\sigma(r) = 3.3 \times 10^{-3}$ when comparing baseline and goal levels. Fig.~\ref{fig:contours_extra_configs} (top) shows the confidence contours and posterior probabilities for three representative masked regions for this instrumental configuration, analogous to Fig.~\ref{fig:contours} (top).

The resulting forecast for the extended frequency configuration on the SO-SAT mask is $\sigma(r) = 3.1\times10^{-3}$, which is slightly better than the constraint achieved with only the six fiducial SO-SAT channels. For this configuration, the spectral parameter $T_{\rm dust}$ was still kept fixed to 19\,K. In Sec.~\ref{sec:using-high-frequency-channels}, we discuss why our $\sigma(r)$ forecast is about the same (or improves a little) even though in this case we are including additional information from three extra frequency channels to constrain the thermal dust. Fig.~\ref{fig:contours_extra_configs} (bottom) shows the confidence contours and posterior probabilities for the same three representative regions for this extended instrumental configuration.

\section{Discussion} \label{sec:discussion} 

\subsection{Adding extra high-frequency channels in CCA} \label{sec:using-high-frequency-channels} 

We have verified that the most important contributor to the polarized foreground contamination in the fiducial six-SO-SAT-frequency channel configuration is the thermal dust. We have estimated the dust and synchrotron residuals separately by mixing simulations that only contain each corresponding foreground with the CCA-estimated weights. In these, the synchrotron residual is significantly smaller than the thermal dust one, so from this point forward, we focus our discussion on the effect of thermal dust. However, a similar analysis can be repeated for synchrotron and how its residual can be mitigated.

The high-frequency channels from CCAT-prime are highly dust dominated, and as such, they can measure the thermal dust precisely. They also have the potential to introduce strong foreground contamination to the estimated CMB signal if we do not model the SED perfectly.  Whether adding such channels gives better or worse results in terms of CMB recovery critically depends on the complexity of the foreground sky in the considered region and the adequacy of the adopted component separation model to represent it. In our case, the balance is not quite so favorable, as the simulated thermal dust has a spatially variable $T_{\rm dust}$ and $\beta_{\rm dust}$, while our model adopts constant values within a given patch.
We note that the small improvement in $\sigma(r)$ when adding the CCAT-prime high-frequency channels is a consequence of dust residuals contaminating the primordial $BB$, which is due to this mismatch between the simulated model and component separation modeling.

First, we have checked that if we consider simpler simulations, e.g. a spatially constant MBB SED, the agreement between the simulation and the component separation modeling is much closer for thermal dust, therefore the thermal dust residuals are smaller and there is a noticeable improvement in $\sigma(r)$ when using the extra information from the high-frequency CCAT-prime channels.

\begin{figure}
    \centering
    \includegraphics[width=1.0\columnwidth]{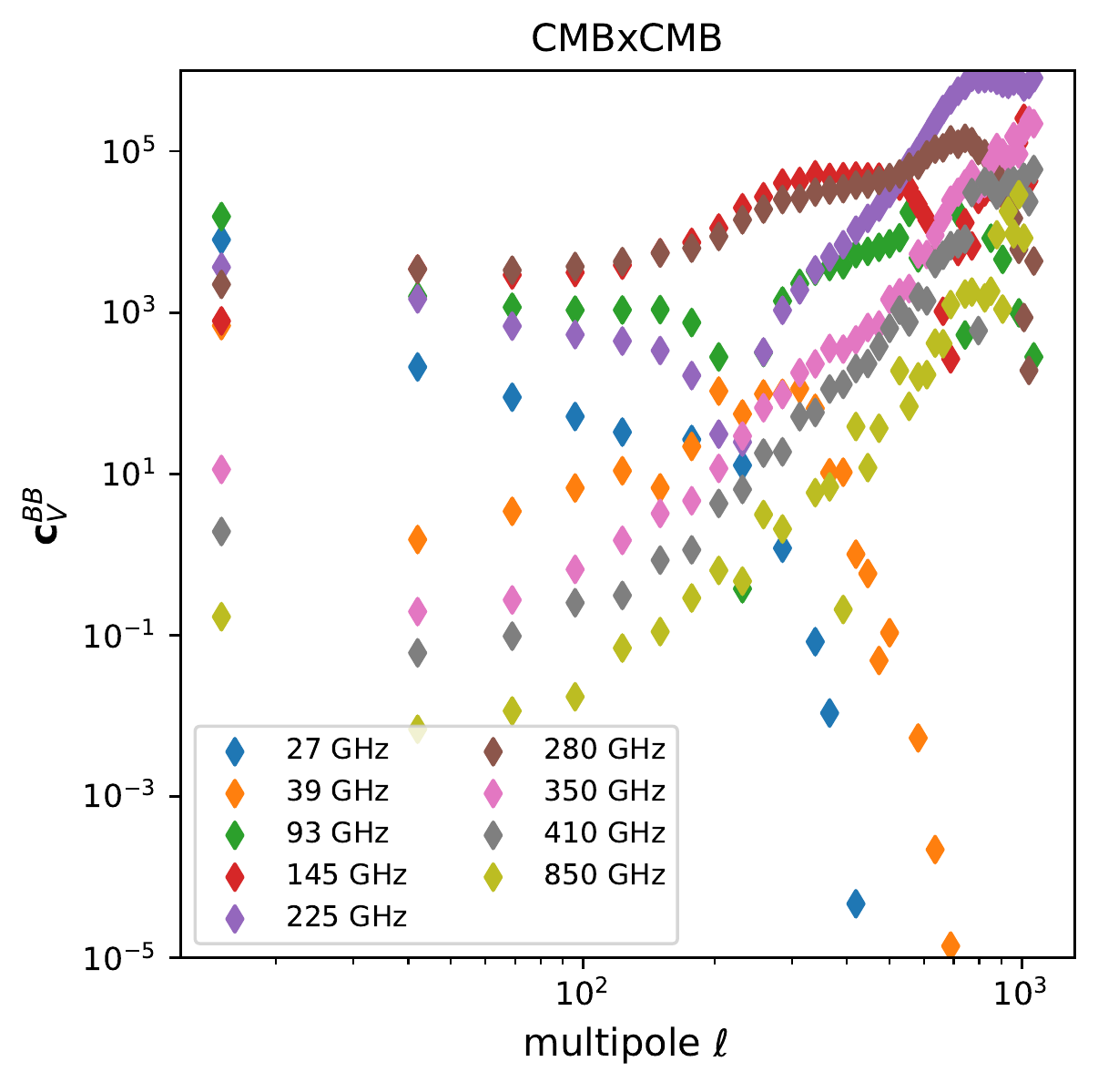}
    \caption{Example of the contribution per frequency channel for the reconstructed CMBxCMB $D_{\ell}^{BB}$ when we run the extended configuration with goal noise levels in the SO-SAT mask. We show the term in eq.~\ref{eq:weights}, which is the weight times the $\mathbf{d}_V$ vector, which corresponds to the power spectra of the full mix minus the power spectra of a noise realization. We only plot the 9 auto-$BB$ power spectra for the 9 frequency channels, we omit the other 36 cross-power spectra for clarity, but the CCA uses all 45 auto- and cross-power spectra in its calculation.}
    \label{fig:weights_example}
\end{figure}

\begin{figure*}
    \centering
    \includegraphics[width=1.0\textwidth]{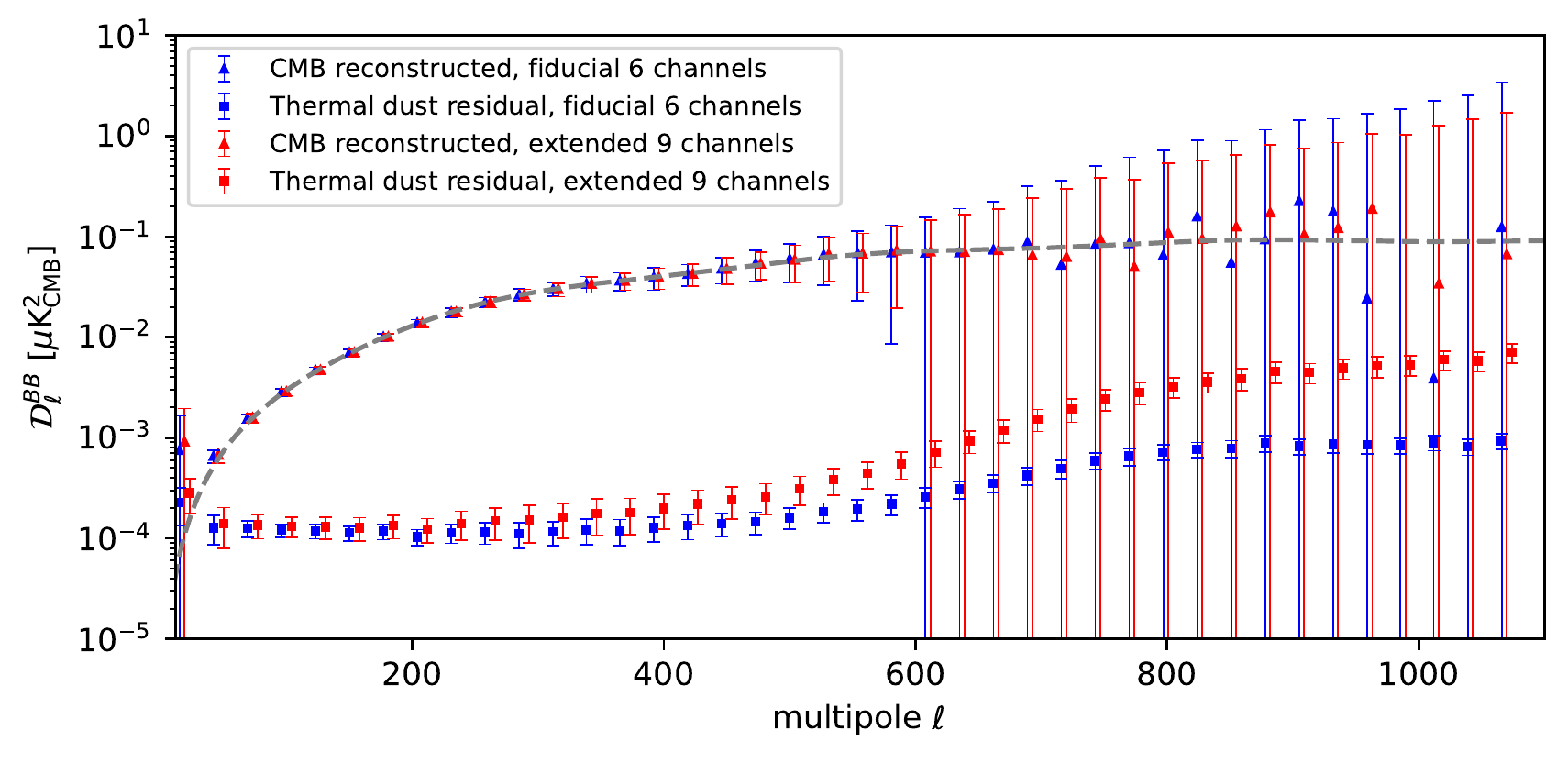}
    \caption{$D_{\ell}^{BB}$ residuals from thermal dust (squares) and reconstructed CMB (triangles) for the SO-SAT mask, for the configurations with goal noise levels, six fiducial channels versus nine extended channels. This shows the mean over the 200 Monte Carlo simulations, while the error bar is the standard deviation over the 200 simulations. The gray line is the fiducial CMB $BB$ spectrum with $r=0$.}
    \label{fig:bb_dust_residuals}
\end{figure*}

To illustrate the issue in our simulation with spatially variable spectral parameters, we can look at the weights that are used to mix the observed data power spectra $\mathbf{d}_V$, described in eq.~\ref{eq:weights}. Fig.~\ref{fig:weights_example} shows the product of the weights and the data vector $\mathbf{d}_V = \tilde{C}_{x}(\hat{\ell}) - \tilde{C}_{n}(\hat{\ell})$ for all multipole bins $\hat{\ell}$, where $\mathbf{d}_V$ is the noise-bias-subtracted full-mix signal. This is for the CMB $D_{\ell}^{BB}$ within the SO-SAT mask with the extended configuration and goal noise levels. For clarity, this figure shows only the contributions for the auto-power spectra for each of the 9 frequency channels, even though the CCA method uses all 45 auto- and cross-spectra between the frequency channels. This figure illustrates the relative contribution from each channel to reconstruct the CMB angular power spectrum $D_{\ell}^{BB}$. For example, the 93, 145, and 225\,GHz channels contribute the most to the CMB, which makes sense because these are the frequencies where the CMB emission is strongest. Conversely, the low and high frequencies of 27, 39, 350, 410, and 850\,GHz have the least relative contribution, which also makes sense. Ideally the CCAT-prime channels weights would be small, but just enough to subtract the dust out of the the channels where the CMB is strong. For the thermal dust, the mismatch in the simulated SEDs and the component separation modeled SEDs in CCA will result in errors in the weights, with the effect of high-frequency channels contaminating the CMB reconstruction with thermal dust residuals. If the weights for the high-frequency channels are small, but they are multiplied by a very strong thermal dust power spectra, such as the one present in the 350, 410, or 850\,GHz channels, they will introduce some dust residual in the CMB reconstruction, which will contaminate the $B$-modes in a significant way.

As another example, we can look at the $BB$ thermal dust residuals directly. In Fig.~\ref{fig:bb_dust_residuals}, we plot the reconstructed CMB $BB$ spectrum as triangles in the SO-SAT mask for the fiducial (blue) versus extended (red) configurations, both with goal noise levels. Shown in the plot is the mean over the 200 Monte Carlo simulations, while the error bar corresponds to the standard deviation of that. We also plot the $BB$ thermal dust residuals (squares) for the same two configurations. These are calculated with the maps of only the simulated thermal dust, mixed with the CCA-calculated weights used for the full simulation for the corresponding case. This is done in the exact same manner as how we calculate $C_{\ell}^{BB,\rm foregrounds}$ in eq.~\ref{eq:model}. Both configurations have roughly the same error bar in their estimation of the CMB, as well as roughly the same dust residual. The nine-channels extended configuration has a slightly smaller nondiagonal term in the Fisher matrix, which means that $r$ and $A_{\rm fore}$ are slightly less correlated, and this configuration has a slight edge over the 6-channel fiducial configuration. We do not see the dust residuals diminishing substantially when adding the extra information from three high-frequency channels, which suggests that our thermal dust component separation modeling is not detailed enough. 

We know that parametric methods might not have enough complexity to model the dust residuals that will bias $r$ \citep{2018ApJ...853..127H}. We can look at examples in the literature that increase the modeling complexity for the thermal dust SED by, for example, using an improved pixel-based foreground parameterization component separation, which allows for the partition of the sky into disjoint patches that can be modeled independently \citep{2019PhRvD..99d3529E}. In recent years the moment expansion has shown promise when performing thermal dust component separation \citep{2017MNRAS.472.1195C,2021A&A...647A..52M,2021MNRAS.503.2478R,2021JCAP...05..047A}. In these methods, the dust spectral parameters are modeled as some value with a small perturbation, so we can account for a higher-order expansion when describing the spatial variability of the dust SED. This is also handy when describing deviations from a perfect MBB spectrum, as we should expect from the measured dust frequency decorrelation \citep{2017A&A...599A..51P,2021ApJ...919...53C}. We can always increase the complexity of the modeling by adding more parameters to the model, but this will increase the degrees of freedom and ultimately dilute the statistical certainty of the estimates. This analysis stresses the importance of careful modeling in the component separation. Dust-monitor channels can be included for improving the dust modeling (e.g. fitting $T_{\rm dust}$), but our analysis suggests that maybe they should not be used naively in the CMB reconstruction itself. We are not saying that using the extra high-frequency information is not useful but rather that our component separation modeling is not good enough in dealing with the residuals that arise by modeling as spatially constant something that is actually spatially variable. Extra care is warranted when modeling components and minimizing the mismatch between observations and models, especially if one wishes to include such high-frequency channels in the component separation analysis. 

\subsection{How Correlated Is $\sigma(r)$ with the foreground Contamination?} Given our analysis, we now attempt to answer the question that we posed at the beginning of this work: Is the region of sky where the foregrounds are at a minimum also the region that yields the smallest possible $\sigma(r)$?

\begin{figure}
    \centering
    \includegraphics[width=1.0\columnwidth]{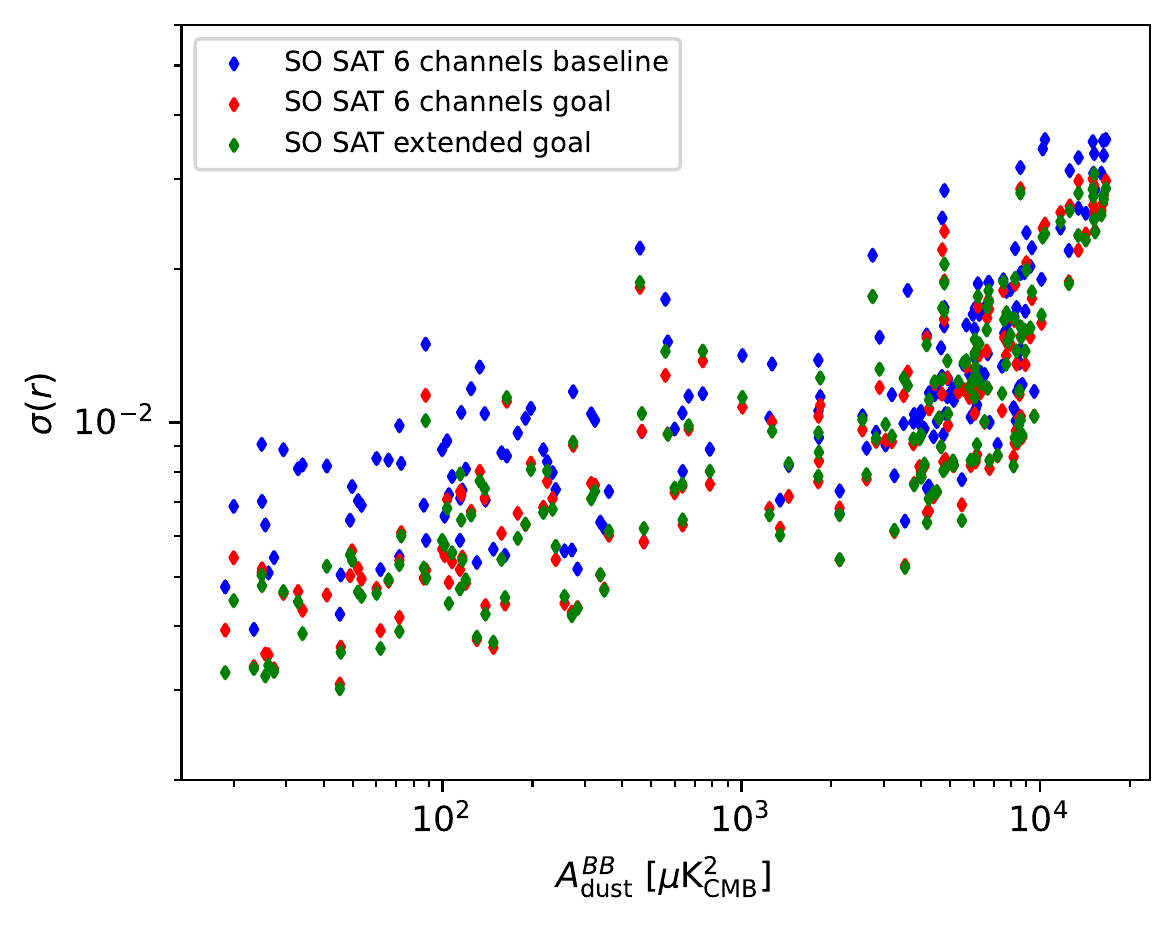}
    \caption{
    Correlation between the $\sigma(r)$ forecast for method A and the amplitude of polarized thermal dust $BB$ power spectrum over the 192 masked regions for the $\sigma(r)$ value after marginalization over $A_{\rm fore}$. The blue points correspond to the fiducial SO-SAT channels with the baseline noise level, the red points are for the same with the goal noise levels, and the green points are the extended configuration with the goal noise levels.
    }
    \label{fig:sigmar_vs_dustpol}
\end{figure}

In Fig.~\ref{fig:sigmar_vs_dustpol}, we plot the $\sigma(r)$ forecast (for method A) versus the polarized intensity of the thermal dust for each of the 192 masked regions we consider in this work. The polarized thermal dust intensity is measured by calculating the $BB$ angular power spectrum of the 2018 data release Planck 353\,GHz frequency map \citep{2020A&A...641A...3P} as prescribed in \citet{planck_2018_xi}, and fitting a power law to it in the multipole range $\ell = 40-599$, given by $\mathcal{D}_{\ell}^{BB} = A_{\rm dust}^{BB} (\ell / 80)^{\alpha_{BB}+2}$. In the figure, we plot the $A_{\rm dust}^{BB}$ amplitude. In blue, we show the fiducial configuration with baseline noise levels, the red points are the same as the goal noise level, and the green points are the extended configuration with the goal noise levels.

This figure shows that when polarized thermal dust is weak (roughly when $A_{\rm dust}^{BB} < 10^2$\,$\mu$K$^2$), the forecasted $\sigma(r)$ is roughly constant, meaning that in the cleanest regions of the sky, we can obtain a similar $BB$ detection performance, regardless of the exact foreground contamination. Then, for regions with stronger foregrounds, $\sigma(r)$ increases slightly with $A_{\rm dust}^{BB}$, until we reach regions where $\sigma(r)$ increases exponentially, roughly where $A_{\rm dust}^{BB}>10^2$\,$\mu$K$^3$. Regarding the difference in $\sigma(r)$ for regions with similar dust contamination, we have checked that this is an effect due to the uncertainty on $A_{\rm fore}$, either through a higher synchrotron contamination or more spatial complexity of the dust spectral parameters in the region.

\section{Conclusions} \label{sec:conclusions}

In our work, we have determined the best regions of the sky to survey for primordial $B$-modes, and we find that the least dust-contaminated regions of the sky, close to the south Galactic pole, have a similar performance. We have performed this forecast for the $\sigma(r)$ error bar for different survey areas over the full sky, using an SO-SAT-like experiment as our example. We consider the fiducial configurations that are scheduled to begin observations in the next couple of years, as well as analyzing the effect of adding the thermal dust-monitor high-frequency channels that will be available with the operation of the CCAT-prime project. In our analysis, we use the CCA method, which works in harmonic space minimizing the observed data covariance, by finding the optimal spectral parameters for the modeled foreground components. We perform our analysis over 200 Monte Carlo simulations and estimate the $\sigma(r)$ error bar by using the Fisher information matrix. We perform forecasts for 192 circular masked regions covering the entire celestial sphere, each equivalent to $f_{\rm sky}=0.1$, and we include the more realistic SO-SAT mask as well.

In our results the SO-SAT survey could measure $r$ with an error bar of $\sigma(r) = 4.9-5.2 \times 10^{-3}$ using the SO-SAT mask with the fiducial six-channel configuration, a baseline noise level, and depending on the method how we account for foreground residuals. When we consider the more sensitive goal noise levels, our forecast improves to $\sigma(r) = 3.3-3.4 \times 10^{-3}$ using the SO-SAT mask. These values are considering a model where we account for the foreground residuals, and marginalizing over this increases the value of $\sigma(r)$ somewhat. However, this paper should not be interpreted as an official SO forecast, as the aim of this work is different: to compare performance for distinct survey regions over the full sky.

We also analyze the potential impact of including three additional high-frequency channels from the CCAT-prime experiment that monitor the strong thermal dust emission. In our analysis, adding these extra high-frequency channels improves the measurement of the CMB $BB$ spectrum slightly. We have investigated the reason for this, and conclude that a naive analysis will leak small dust residuals from these extra channels that have a very strong dust component. Even if the polarized dust residuals are very small, they are still quite significant for $B$-modes. Any mismatch in the thermal dust SED when we model it in our component separation will create residuals that will affect the reconstruction of the CMB $B$-mode spectrum. This analysis emphasizes the need to accurately model the SED of the CMB foregrounds, especially for the thermal dust polarization.

We attempt to answer the question posed at the beginning of this work, that is, to measure $r$, is it better to use the region that is cleanest of foregrounds, or is there a sweet spot where the foreground amplitudes are optimal from the point of view of facilitating their removal? We know the most important factor for mitigating foreground residuals is to model the SED correctly. From our analysis, there appears to be a range of sky regions where the foreground contamination is weak and the $\sigma(r)$ error bar is roughly the same. This seems to support the notion that having the smallest possible amount of foregrounds and having a slightly higher amount of foreground will yield a similar measurement of $r$, meaning that component separation is doing its job by mitigating $BB$ foreground contamination in a range. One does not need the absolute best possible region for a $B$-mode survey to yield a good result. Nevertheless, there is a limit to this, and for regions with higher foreground contamination, the achievable $\sigma(r)$ increases.

We conclude by stressing the importance of characterizing the SED for polarized foregrounds in the component separation analysis, in particular for polarized thermal dust residuals, which will significantly bias a measurement of $B$-modes if not accounted for. From our results, we can conclude that the cleanest regions of the sky will yield the overall best results in terms of measuring the primordial $B$-modes. There is some tolerance for regions with a slightly higher polarized foreground contamination, because the component separation is able to fulfill its purpose. Our goal is to mitigate the polarized dust residual as much as possible, and to do this, we need to improve the modeling of polarized foregrounds SEDs in the component separation. In parallel, the instrumentation that improves the sensitivity is another important factor. The need to better understand foregrounds and their SED is imperative, and new experiments that will monitor ancillary frequencies for synchrotron and thermal dust, such as QUIJOTE \citep{2015MNRAS.452.4169G}, C-BASS \citep{2018MNRAS.480.3224J}, and the already-mentioned CCAT-prime will be vital in the endeavor to detect primordial $B$-modes.

\acknowledgements{
C.H.C. acknowledges the funding from Becas Chile/CONICYT. C.H.C. and K.M.H. are supported by NASA through ATP award NNX17AF87G and by NSF through AAG awards 1815887 and 2009870. We thank Stefano Camera, David Alonso, Christian Reichardt, and Michael D. Niemack for useful suggestions that improved this work. We thank the anonymous referee for their useful comments. This work uses public data from Simons Observatory. Some of the computing for this project was performed on the HPC cluster at the Research Computing Center at the Florida State University (FSU). 
}
\software{ \textsc{healpix} \citep{2005ApJ...622..759G}, 
\textsc{namaster} \citep{2019MNRAS.484.4127A},
\textsc{pysm3} \citep{thorne_2017},
\textsc{camb} \citep{2012JCAP...04..027H}
}

\bibliography{biblio}{}
\bibliographystyle{aasjournal}

\end{document}